\documentclass{aastex631}
\usepackage{hyperref}
\usepackage{amsmath}
\newcommand{\pd}[2]{\frac{\partial #1}{\partial #2}}
\newcommand{\dd}[2]{\frac{d #1}{d #2}}
\newcommand{\floor}[1]{\lfloor #1 \rfloor}

\newcommand{\paren}[1]{\left( #1 \right)}

\usepackage[utf8]{inputenc}
\usepackage{float}
\usepackage{natbib}

\begin{document}
\title{Action-Angle Variables for Axisymmetric Potentials via Birkhoff Normalization}
\date{April 2024}
\author[0000-0002-1032-0783]{Sam Hadden}
\affiliation{Canadian Institute for Theoretical Astrophysics, 60 St George St Toronto, ON M5S 3H8, Canada}

\begin{abstract}
     We describe a method for calculating action-angle variables in axisymmetric galactic potentials using Birkhoff normalization, a technique from Hamiltonian perturbation theory. An advantageous feature of this method is that it yields explicit series expressions for both the forward and inverse transformations between the action-angle variables and position-velocity data. It also provides explicit expressions for the Hamiltonian and dynamical frequencies as functions of the action variables. We test this method by examining orbits in a Miyamoto-Nagai model potential and compare it to the popular Stäckel approximation method. When vertical actions are not too large, the Birkhoff normalization method achieves fractional errors smaller than a part in $10^{3}$ and outperforms the Stäckel approximation. 
     We also show that the range over which Birkhoff normalization provides accurate results can be  extended by constructing Padé approximants from the perturbative series expressions developed with the method.  Numerical routines in \texttt{Python} for carrying out the Birkhoff normalization procedure are made available.
\end{abstract}
\section{Introduction}
While a comprehensive dynamical description of a galaxy must take into account mutual gravitational interactions among stars, dark matter, and gas, dramatically simplified models that consider the orbits of stars in static, smooth potentials provide surprisingly powerful tools for understanding the dynamics of these complex systems. 
Numerical experiments show that typical  stellar orbits in model galactic potentials execute quasi-periodic motions \citep[e.g.,][]{binney_spectral_1982,binney_spectral_1984}. Such motions are restricted to 3-dimensional tori embedded in the full 6-dimensional phase space spanned by stars' spatial positions and velocities. For completely integrable Hamiltonian systems,  action-angle (AA) variables provide a particularly convenient set of canonical coordinates for describing motion on such tori \citep[e.g.,][]{Goldstein2002}.  Fixing the values of  three action variables, 
$\pmb{J}$, specifies an individual torus as a 3-dimensional submanifold of the 6-dimensional phase space, while the canonically conjugate angle variables, $\pmb{\theta}$, serve as coordinates on the torus. Furthermore, the time evolution of trajectories on a given torus are described simply by 
the linear advance of the angle variables with time at rates $\frac{d}{dt}\pmb{\theta} = \nabla_{\pmb{J}}H$, where $H$ is the Hamiltonian function.

While AA variables are a powerful theoretical tool, it is seldom possible to compute them explicitly from stellar position and velocity data for potentials commonly adopted when modeling our Milky Way and other galaxies.  This is because these potentials generally do not admit a set of three globally-defined isolating integral which are required in order to have a well-defined transformation to AA variables everywhere in phase space. Nonetheless, the quasi-periodic nature of stellar orbits over much of the phase space of interest in galactic potentials makes its possible to construct local transformations to AA variables via numerical means.

Numerous numerical methods have been developed for the purposes of constructing AA variables in galactic potentials \citep[see review by][]{sanders_review_2016}. One of the most widely-used methods is the Stäckel approximation or ``Stäckel fudge", originally proposed by \citet{binney_actions_2012}.
In axisymmetric potentials, this method allows radial and vertical actions to be estimated via numerical quadratures by implicitly treating the underlying potential as though it were in Stäckel form, meaning separable in prolate spheroidal coordinates.
While the Stäckel approximation usually provides quite accurate determinations of AA variables, it does have some shortcomings. First, the transformation cannot be readily inverted to provide positions and velocities as functions of actions and angle nor express the original Hamiltonian in terms of AA variables. Furthermore, because the method requires evaluation of integrals by numerical quadratures, it can become computationally expensive when evaluating transformations for large amounts of data. 

In this paper, we present a perturbative method to determine action-angle variables in axisymmetric galactic potentials. 
This method is based on a method originally proposed by \citet{birkhoff1927dynamical} for developing successively refined approximations to the motion of Hamiltonian systems in the vicinity of elliptic equilibria. 
It is variously referred to as to as ``Birkhoff normalization" \citep[e.g.,][]{Deprit1969} or the ``Birkhoff-Gustavson method" \citep[e.g.,][]{lowenstein2012essentials}.
One of the principal advantages of this method is that it provides explicit expressions for the transformation from position and velocity data to AA variables. 
Furthermore, this method simultaneously yields explicit expressions for the inverse transformation, meaning stellar positions and velocities as well as the Hamiltonian can be represented explicitly in terms of the AA variables. Expressions are constructed as multi-variate polynomials in complex canonical coordinates, introduced in Section \ref{sec:algorithm}, and therefore can be evaluated at minimal computational cost. 
Together, these features can make Birkhoff normalization an attractive alternative to the Stäckel approximation, depending on the application. 

The use of perturbation theory in galactic dynamics to derive approximate integrals of motion is not new \citep[e.g.,][]{contopoulos_existence_1963,saaf_formal_1968,de_zeeuw_stellar_1983}.
In fact, \citet{gustavson_constructing_1966} implements and applies an algorithm  very similar to the one presented here to derive a formal second integral of the Hénon–Heiles potential \citep{henon_applicability_1964}.
The current paper builds on this past work in at least a couple respects. First, we follow \citet{Deprit1969} and develop perturbative series using the Lie transform method. 
This method allows any functions defined on phase space in terms of the original canonical variables to be readily expressed in transformed variables and vice versa.
This fact is exploited  to construct an explicit expression for the canonical angle variable, $\theta_\phi$, associated with orbits' azimuthal degree of freedom, thereby providing a complete transformation from 6D position and velocity data to AA variables. 
Second, we show how Padé approximants can be deployed to extend the range of validity of the series expressions furnished by the method. 
Finally, a modern computational implementation of the method in \texttt{Python} is made publicly available.\footnote{\href{https://github.com/shadden/AA_variables_via_Birkhoff_Normalization}{github.com/shadden/AA\_variables\_via\_Birkhoff\_Normalization}}
The availability of computer algebra and automatic differentiation packages in modern languages like \texttt{Python}  make it feasible to apply Birkhoff normalization to more realistic potential models than the simple  Hénon–Heiles model originally considered by \citet{gustavson_constructing_1966}.

The plan of this paper is as follows: we introduce the algorithm in Section \ref{sec:algorithm}. In Section \ref{sec:applications}, we demonstrate the method with an example application to stellar orbits in  Miyamoto-Nagai potential \citep{MiyamotoNagai1975}, a classic model for the potential of a galactic disk. Finally, we summarize and discuss our results in  Section \ref{sec:summary}.

\section{The Algorithm}
\label{sec:algorithm}
\subsection{Setting}
  We consider orbits in an axisymmetric potential, $\Phi$. The Hamiltonian, expressed in terms of polar coordinates $(R,z,\phi)$ along with their canonically conjugate momenta, $(p_R,p_z,L)$, is
 \begin{equation}
     \mathcal{H}(R,z,\phi,p_R,p_z,L) = \frac{1}{2}\left(p_R^2 + p_z^2 + \frac{L^2}{R^2}\right) + \Phi(R,z)~.\label{eq:h_full}
 \end{equation}
Since the Hamiltonian in \eqref{eq:h_full} has no explicit $\phi$ dependence, the angular momentum, $L$, is conserved and we can consider the dynamics in the reduced phase space comprised of $(z,R,p_z,p_R)$, treating $L$ as a parameter. 
The Hamiltonian governing the dynamics in this reduced phase space is then given by 
\begin{equation}
        {H}(R,z,\phi,p_R,p_z;L) = \frac{1}{2}\left(p_R^2 + p_z^2\right) + {\Phi}_\mathrm{eff}(R,z;L)\label{eq:h_reduced}
\end{equation}
where  we have defined 
    ${\Phi}_\mathrm{eff} ={\Phi}({R,z}) + \frac{L^2}{2R^2}$ 
as the effective potential.
For a given value of angular momentum, $L$, there is a (typically unique) circular orbit of radius of $R_C(L)$ which satisfies 
\begin{equation}
\pd{}{R}{\Phi}_\mathrm{eff}({R,z})\bigg|_{(R,z)=(R_C,0)}=0~.
\end{equation}
For nearly circular and planar orbits, the Hamiltonian \eqref{eq:h_reduced} is given approximately by 
\begin{equation}
 H\approx H_2(z, R,p_R,p_z;L) := 
    \frac{1}{2}\left(p_R^2 + \kappa^2\delta R^2\right) 
    +
    \frac{1}{2}\left(p_z^2 + \nu^2 z^2\right)~.
    \label{eq:epicyclic_hamiltonian}
\end{equation}
where $\kappa^2 = \pd{^2}{R^2}{\Phi}_\mathrm{eff}$ and $\nu^2 = \pd{^2}{z^2}{\Phi}_\mathrm{eff}$, with derivatives evaluated at $(R,z) = (R_C,0)$, are the epicyclic and vertical frequencies, respectively and $\delta R= R-R_C$.
Equation \eqref{eq:epicyclic_hamiltonian} is, of course, the Hamiltonian of two un-coupled harmonic oscillators of frequency $\kappa$ and $\nu$ and the equations of motion derived from it provide the so-called ``epicyclic approximation" for stars' orbits \citep[e.g.,][]{BT2008}. 

\subsection{Complex canonical variables}
\label{sec:cc_vars}
The equations of motion derived from Equation \eqref{eq:epicyclic_hamiltonian} are completely integrable and we can perform a canonical transformation, $T : (R,z,p_R,p_z) \mapsto (\phi_R,\phi_z,I_R,I_z)$, to new variables such that $H_2 \circ T^{-1} = \kappa I_R + \nu I_z$.
Throughout this paper, we will frequently work instead with the associated complex canonical variables 
\begin{align}
         x_R &= \sqrt{\frac{\kappa}{2}}\left(\delta R + \mathrm{i}\frac{p_R}{\kappa} \right) 
         = \sqrt{I_R}e^{-\mathrm{i}\phi_R}
         \label{eq:xR_def}
         \\
     x_z &= \sqrt{\frac{\nu}{2}}\left(z + \mathrm{i}\frac{p_z}{\nu} \right) =  \sqrt{I_z}e^{-\mathrm{i}\phi_z}~
     \label{eq:xz_def}
\end{align} 
and their complex conjugates, $\bar{x}_R$ and $\bar{x}_z$.
The Poisson bracket of two functions, $f$ and $g$, of phase space coordinates is expressed in terms of these variables by
\begin{equation}
         [f,g] = -\mathrm{i}\sum_{j}
\left(
    \pd{f}{{x}_j}\pd{g}{\bar{x}_j} 
    - 
    \pd{f}{\bar{x}_j}\pd{g}{{x}_j}
\right)
\label{eq:complex_bracket}
\end{equation}
where the subscript $j$ denotes either $R$ or $z$.
In terms of the complex canonical variables,\footnote{
Here and elsewhere we will make a slight abuse of notation and use the same symbol to denote the Hamiltonian function expressed in terms of complex variables as we do for the Hamiltonian expressed in terms of polar coordinates.} Equation \eqref{eq:epicyclic_hamiltonian} becomes $H_2 = \kappa x_R\bar{x}_R + \nu x_z\bar{x}_z $ and Hamilton's equations give 
$\frac{d}{dt}x_R = [x_R,H_2] = -\mathrm{i}\kappa x_R$ and $\frac{d}{dt}x_z = [x_R,H_2] = -\mathrm{i}\nu x_z$.
Thus, under the epicyclic approximation, the complex canonical variables evolve at constant magnitude with a uniformly rotating phase.
 This will generally no longer be true when higher order terms are included in the Hamiltonian. 
Our goal will be to develop an algorithm that constructs a canonical transformation to new complex canonical variables, $x_R'$ and $x'_z$, for which this is again true up to some specified order in the new variables and their complex conjugates.

\subsection{An illustrative example}
\label{sec:pendulum_example}
Before giving the general algorithm, we first demonstrate its basic principles by way of an  example application to the pendulum. This example provides a pedagogical introduction to Birkhoff normalization and the construction of canonical transformations via the Lie series method with complex canonical variables. The example illustrates the basic ideas behind the method while avoiding some of the cumbersome mathematical notation and manipulations required to formulate the procedure in a more general context. Readers uninterested in the technical details involved in an algorithmic implementation of the Birkhoff normalization procedure can therefore read this section while skipping or skimming Section \ref{sec:algorithm:general}.

 We start with  the Hamiltonian governing the dynamics of a simple pendulum, 
\begin{equation}
    h(\theta,p) = \frac{p^2}{2} - \omega_0^2\cos{\theta}.
    \label{eq:h_pend_pq}
\end{equation}
For small $\theta$, the truncated Hamiltonian, $h_{2}(\theta,p) = (p^2 + \omega_0^2\theta^2)/2$, gives a harmonic oscillator approximation of the dynamics.
In terms of the complex canonical variable $x = \sqrt{\frac{\omega_0}{2}}(\theta + \mathrm{i} p/\omega_0)$, the Hamiltonian  \eqref{eq:h_pend_pq} becomes
\begin{equation}
\label{eq:ham_pend_expansion}
    h(x,\bar{x}) = \omega_0 x\bar{x} + \sum_{n=2}^{\infty}h_{2n}(x,\bar{x})
\end{equation}
where
\begin{equation}
\label{eq:ham_pend_pert_term}
    h_{2n} = 
    \frac{(-1)^{n+1}\omega_0^2}{(2n)!}
    \left(\frac{1}{2\omega_0}\right)^n
    \left(x + \bar{x}\right)^{2n}~.
\end{equation}

We will seek a canonical transformation to a new complex canonical variable, $x'$, so that the transformed Hamiltonian, $h'$, depends only on $|x'|^2$ up to a specified power, $N$, in the transformed variable, $x'$,  and its complex conjugate, $\bar{x}'$.  

The Lie series method \citep[e.g.,][]{ferraz2007canonical,lichtenberg1992} provides a particularly convenient means of constructing such a transformation. This method involves the construction of a generating function, $\chi$, such that the transformation from the new variables to the old variables is given by the time-1 flow of the Hamiltonian vector field generated by $\chi$. 
The time derivative of a function, $f$, under this flow is given by the Poisson bracket $[f,\chi]$ and we define the differential operator $\mathcal{L}_\chi := [\cdot,\chi]$, referred to as the Lie derivative.
This allows us to express a function, $f$, given in terms of the original phase space coordinates $(x,\bar{x})$, in terms of the new phase space variables, $(x',\bar{x}')$, formally as 
\begin{equation}
    f(x,\bar{x}) = (\exp[\mathcal{L}_\chi]f)(x',\bar{x}')  = \paren{\sum_{n=0}^\infty \frac{1}{n!}\mathcal{L}_\chi^nf}(x',\bar{x}')~.
    \label{eq:lie_series_def}
\end{equation}

Returning to our pendulum example, we seek a generating function $\chi$ such that  $h'(x',\bar{x}') = (\exp[\mathcal{L}_\chi] h)(x',\bar{x}') = \sum_{n=1}^{N/2}h'_{2n}(|x'|^2)  + R_{N}(x',\bar{x}')$ where the remainder terms $R_{N}(x',\bar{x'}) \sim \mathcal{O}(|x'|^{N+1})$.
Our approach  will be to write our generating function as $\chi = \sum_{n=2}\chi_{2n}$ where $\chi_{2n} \sim {\mathcal{O}(|x'|^{2n})}$. This will allow us to develop an iterative procedure for determining $\chi_{2n}$ and $h'_{2n}$ at each stage in terms of the functions $h_{2},h_{4}...,h_{2n}$ defined in Equation \eqref{eq:ham_pend_pert_term} and the functions $\chi_{4},\chi_{6},...,\chi_{2n-2}$ determined in previous stages.  Writing out the transformation Equation \eqref{eq:lie_series_def} explicitly in terms of Poisson brackets, we obtain
\begin{align}
    \exp[\mathcal{L}_\chi]h &=
    h + [h,\chi] + \frac{1}{2}[[h,\chi],\chi] + ...
    \\&=
    \underbrace{[h_2,\chi_4] + h_4}_{h'_4}
    +\underbrace{[h_2,\chi_6] + [h_4,\chi_4] +\frac{1}{2}[[h_2,\chi_4],\chi_4]  + h_6}_{h'_6}
    + \mathcal{O}(|x'|^{8}) 
    \label{eq:lie_transform}
\end{align}
where, in the second line, we have grouped terms by common powers of $|x'|$.
Consider the equation
\begin{equation}
 h'_4 = 
 [h_2,\chi_4]
 - 
 \paren{
  \frac{x'^4}{96} + \frac{x'^3 \bar{x}'}{24}+\frac{x'^2 \bar{x}'^2}{16} +\frac{x' \bar{x}'^3}{24}+\frac{\bar{x}'^4}{96}
    }\label{eq:h4prime}
\end{equation}
defining the fourth order term in the transformed Hamiltonian, where we have used Equation \eqref{eq:ham_pend_pert_term} to write out $h_4$ explicitly. We wish to define $\chi_4$ so that $h'_4$ depends only on $|x'|^2$.  Notice that the Poisson bracket between $h_2$ and any monomial, $x'^{k}\bar{x}'^{\bar{k}}$, gives $[h_2,x'^{k}\bar{x'}^{\bar{k}}] = \mathrm{i}\omega_0(k-\bar{k})x'^{k}\bar{x}'^{\bar{k}}$. 
Therefore, if we choose $\chi_4$ as the sum of monomial terms
\begin{eqnarray}
    \chi_4 =
  \frac{1}{\mathrm{i}\omega_0}\paren{\frac{x'^3 \bar{x}'}{48}-\frac{x' \bar{x}'^3}{48}-\frac{\bar{x}'^4}{384}+\frac{x'^4}{384}}
\end{eqnarray}
then each term in the polynomial $[h_2,\chi_4]$ cancels a corresponding term appearing in Equation \eqref{eq:h4prime} except the term $\propto x'^2\bar{x}'^2$.  We therefore have
\begin{equation}
    h'_4 = -\frac{1}{16}x'^2 \bar{x}'^2~,
\end{equation}
which depends only on $|x'|^2$ as desired.

With $\chi_4$ determine, we can write $h'_6$ according to Equation \eqref{eq:lie_transform} as 
\begin{equation}
    h'_6 = [h_2,\chi_6] +\frac{x'^6}{2560 \omega_0 } -\frac{x'^5 \bar{x}'}{3840 \omega_0 }-\frac{7 x'^4 \bar{x}'^2}{1536 \omega_0 }-\frac{x'^3 \bar{x}'^3}{256 \omega_0 }-\frac{7 x'^2 \bar{x}'^4}{1536 \omega_0 }-\frac{x' \bar{x}'^5}{3840 \omega_0 }+\frac{\bar{x}'^6}{2560 \omega_0 }~.\label{eq:h6prime}
\end{equation}
We will again choose $\chi_6$ so that the Poisson bracket $[h_2,\chi_6]$ cancels all of the monomial terms appearing on the right-hand side of Equation \eqref{eq:h6prime} except the term $\propto x^3\bar{x}^3$. Proceeding, we find
\begin{align}
    \chi_6 &=\frac{1}{\mathrm{i}\omega_0^2}\paren{-\frac{x'^6}{15360}+\frac{x'^5 \bar{x}'}{15360}+\frac{7 x'^4 \bar{x}'^2}{3072}-\frac{7 x'^2 \bar{x}'^4}{3072}-\frac{x' \bar{x}'^5}{15360}+\frac{\bar{x}'^6}{15360}}~.
    \\
    h'_6  &= -\frac{x'^3 \bar{x}'^3}{ 256\omega_0 }~.
\end{align}

If we carried on with the expansion of Equation \eqref{eq:lie_transform} to higher order, we would find that at each step,
we need to solve an equation of the form 
\begin{equation}
 h'_{2n} = [h_2,\chi_{2n}] + \Psi_{2n}\label{eq:homological_eq}
\end{equation}
 where $\Psi_{2n}$ is a collection of terms of order $2n$ involving Poisson brackets of the known functions $h_4,h_6,...,h_{2n}$ and $\chi_4,...,\chi_{2n-2}$. In the next section we will detail an algorithm for iteratively constructing and solving \emph{homological equations} like Equation \eqref{eq:homological_eq}. 
 
 Let us conclude our present example by approximating the pendulum's oscillation frequency as a function of maximum angular displacement, $\theta_\mathrm{max}$, in order to illustrate how functions of the new, transformed variables, $x'$ and $\bar{x}'$, may be calculated explicitly from the original coordinates and vice versa. 
Our transformed Hamiltonian is given by 
\begin{equation}
    h' = \omega_0 |x'|^2 - \frac{1}{16}|x'|^4  - \frac{1}{256 \omega_0 }|x'|^6 + \mathcal{O}(|x'|^8)
\end{equation}
so the oscillation frequency will be given, in terms of the action $J= |x'|^2$, by $\omega(J) = \omega_0 - \frac{1}{8}J - \frac{3}{256\omega_0}J^2$. We therefore need the value of $J$ in terms of $\theta_\mathrm{max}$.

Given any function, $f$, on the phase space expressed in terms of the transformed variables, $x'$ and $\bar{x}'$, its expression in terms of the original, un-primed variables $x$ and $\bar{x}$ is given by  
\begin{equation}
\label{eq:inverse_transform}
   f\left(x'(x,\bar{x}),\bar{x}'(x,\bar{x})\right) = (\exp[-\mathcal{L}_\chi]f)(x,\bar{x})
\end{equation}
since $\exp[-\mathcal{L}_\chi]f$ gives the function $f$ evaluated at the time-($-1$) flow of the Hamiltonian vector field generated by $\chi$.
Applying Equation \eqref{eq:inverse_transform} to the function $\omega : (x',\bar{x}')\mapsto \omega_0 - \frac{1}{8}|x'|^2 - \frac{3}{256\omega_0}|x'|^4$, we obtain  the oscillation frequency in terms of the original, un-transformed complex variable $x = \sqrt{\frac{\omega_0}{2}}(\theta + \mathrm{i}p/\omega_0)$ as
\begin{align}
\label{eq:omega_vs_xxbar}
    (\exp[-\mathcal{L}_\chi]\omega)(x,\bar{x}) & = 
    \omega_0   - \frac{|x|^2}{8}
     -\frac{3}{256}|x|^4 + \frac{1}{8}\left[|x|^2,\chi_4\right]+\mathcal{O}(|x|^6)\\
     &=\omega_0   - \frac{x\bar{x}}{8} + \frac{1}{\omega_0}\paren{\frac{\bar{x}^{4}}{768} + \frac{\bar{x}^{3} x}{192} - \frac{3 \bar{x}^{2} x^{2}}{256} + \frac{\bar{x} x^{3}}{192} + \frac{x^{4}}{768}} + \mathcal{O}(|x|^6)~.
\end{align}
 Substituting $x = \sqrt{\frac{\omega_0}{2}}\theta_\mathrm{max}$ into Equation \eqref{eq:omega_vs_xxbar} gives
\begin{equation}
     \omega \approx \omega_0\paren{1 -\frac{\theta _{\max }^2}{16} + \frac{\theta _{\max }^4}{3072}+...},
\end{equation}
which matches the first three terms of the Taylor expansion of the exact result, $\omega = \frac{\pi  \omega_0 }{2 \mathbb{K}\left(\sin ^2\left(\frac{\theta_{\max } }{2}\right)\right)}$, where $\mathbb{K}$ denotes the complete elliptic integral of the first kind \citep[e.g.,][]{lichtenberg1992}.
\subsection{The general algorithm}
\label{sec:algorithm:general}
The Birkhoff normalization procedure, illustrated above in the case of the pendulum, readily generalizes to  systems with $d>1$ degrees of freedom. 
Specifically, the procedure can be applied in the vicinity of an elliptic equilibrium to construct a formal series for a generating function, $\chi$, that transforms the Hamiltonian so that it only depends on $d$ action-like variables. While these formal series are, in general, divergent, their finite-order truncations are nonetheless often very useful for approximating a given system's dynamics \citep[e.g.,][]{contopoulos_non-convergence_2003,efthymiopoulos_nonconvergence_2004}. 
The numerical results presented in later in Section \ref{sec:applications} 
show that the approximate integrals  computed for our chosen order of truncation are very nearly constants over a large region of the phase space.

We apply the Birkhoff normalization procedure to orbits in the vicinity of  planar, circular orbits of axisymmetric galactic potentials, which are elliptic equilibria of the two degree-of-freedom Hamiltonian given in Equation \eqref{eq:h_reduced}.  In terms of the complex canonical variables  $x_R$ and $x_z$ introduced in Section \ref{sec:cc_vars}, the Hamiltonian can be written as
\begin{eqnarray}
    H = \kappa x_R\bar{x}_R + \nu \kappa x_z\bar{x}_z + \sum_{n=3}^{\infty}H_n(x_R,x_z,\bar{x}_R,\bar{x}_z)
    \label{eq:H_grouped}
\end{eqnarray}
    where
\begin{eqnarray}
    H_n(x_R,x_z,\bar{x}_R,\bar{x}_z) = 
    \sum_{m=0}^n
    \frac{\Phi_\mathrm{eff}^{(m,n-m)}}{(n-m)!m!}
    \paren{\sqrt{\frac{1}{2\kappa}}}^{n-m}
    \paren{\sqrt{\frac{1}{2\nu}}}^{m}
    \sum_{p=0}^{m}
    \sum_{q=0}^{n-m}
    \binom{m}{p}
    \binom{n-m}{q}
    x_R^{p}
    x_z^{q}
    \bar{x}_R^{m-p}
    \bar{x}_z^{n-m-q}
    \label{eq:H_n_term}
\end{eqnarray}
and  $\Phi_\mathrm{eff}^{(m,n-m)} = \pd{^n}{R^{m}z^{n-m}}\Phi_\mathrm{eff}\big|_{(R,z)=(R_C,0)}$.  Values of the partial derivatives of the effective potential, $\Phi_\mathrm{eff}^{(m,n-m)}$,  up to the desired maximum order of the perturbative expansion are required to carry out the Birkhoff normalization procedure. Calculation of these derivatives is most conveniently done with the aid of a computer algebra system or automatic differentiation packages.

Equation \eqref{eq:H_grouped} represents the Hamiltonian as a series of terms grouped by their order in the complex canonical variables. We will construct similar series for our generating function, $\chi = \sum_{k=k_\mathrm{min}}^\infty \chi_{k}$, and the transformed Hamiltonian, $H'=\sum_{m=1}^\infty H'_{2m}$ through an iterative process.  Note that because $H'$  depends only on the action variables, it only contains even-order terms.

\subsubsection{Grouping Lie Series Terms by Order}
To construct our series solution, it will be necessary to group terms of the same order in the expansion of the Lie series $H' = \exp[\mathcal{L}_\chi] H$.  This is most readily accomplished using recursion formulae originally derived by \citet{Deprit1969}. Below, we give a derivation of these formulae that closely follows the presentation of \citet[][Chapter 5]{ferraz2007canonical}.

If $f_k$ and $g_l$  are homogeneous polynomials of degree $k$ and $l$, respectively, in complex canonical variables then $[f_k,g_l]$ is a homogeneous polynomial of degree $k+l-2$ in those variables.
Define  $D_{k} := [\cdot ,\chi_k ]$ so that,
given series expansions $f=\sum_{k'=k'_\mathrm{min}}^\infty f_{k'}$, and $\chi = \sum_{k=k_\mathrm{min}}^\infty \chi_{k}$, we can write
\begin{align}
\mathcal{L}_\chi f &= 
    \sum_{k=k_\mathrm{min}}^\infty \sum_{k'=k'_\mathrm{min}}^\infty D_{k}f_{k'} 
    =
    \sum_{l=k_\mathrm{min} + k'_\mathrm{min}-2}^\infty
    \sum_{k=k_\mathrm{min}}^{l+2-k'_\mathrm{min}}D_{k}f_{l+2-k} 
    :=
    \sum_{l=l_\mathrm{min}^{(1)}}^\infty\Upsilon^{1}_{l}(f,\chi) \nonumber\\
    \label{eq:Df_defn}
\end{align}
where $l_\mathrm{min}^{(1)} = k_\mathrm{min} + k'_\mathrm{min} - 2$ and $\Upsilon^{1}_{l}(f,\chi)$ represents all terms of order $l$ appearing in an expansion of $\mathcal{L}_{\chi}f$. If we now substitute $\mathcal{L}_{\chi}f$ for $f$ in Equation \eqref{eq:Df_defn},
we obtain
\begin{equation}
 \mathcal{L}^2_{\chi}f =  \sum_{l=k_\mathrm{min}+l_\mathrm{min}^{(1)}-2}^\infty
    \sum_{k=k_\mathrm{min}}^{l+2-l_\mathrm{min}^{(1)}}D_{k}\Upsilon^1_{l+2-k}(f,\chi) :=
    \sum_{l=l_\mathrm{min}^{(2)}}^\infty\Upsilon^{2}_{l}(f,\chi)
\end{equation}
where $l_\mathrm{min}^{(2)} = k_\mathrm{min} + l^{(1)}_\mathrm{min} - 2$.
More generally, we have
\begin{align}
l_\mathrm{min}^{(n)} &= k_\mathrm{min} + l^{(n-1)}_\mathrm{min} - 2\\
 \mathcal{L}^{n}_{\chi}f &= \sum_{ l=l_\mathrm{min}^{(n)} }^\infty 
          \sum_{k=k_\mathrm{min}}^{l+2-l_\mathrm{min}^\mathrm{(n-1)}}D_k\Upsilon^{(n-1)}_{l+2-k}(f,\chi)
          :=\sum_{ l=l_\mathrm{min}^{(n)} }^\infty \Upsilon^{n}_{l}(f,\chi)
\end{align}
In other words,  $\Upsilon^{n}_l$ is the collection of order $l$ terms arising from the $n$th Lie derivative in our expansion. Values of $\Upsilon^{n}_l$ for a given $n$ and $l$ can be computed using the  recursion formulae 
\begin{align}
\Upsilon^{0}_{l}(f,\chi) &= f_{l}\\
l_\mathrm{min}^{(0)} &= k'_\mathrm{min}\\
\Upsilon^{n+1}_{l}(f,\chi) &= \sum_{k=k_\mathrm{min}}^{l+2 - l^{(n)}_\mathrm{min}} D_k\Upsilon^{n}_{l+2-k}(f,\chi)\\
l_\mathrm{min}^{(n+1)} &= k_\mathrm{min} + l^{(n)}_\mathrm{min} - 2~.
\label{eq:lmin}
\end{align}
\subsubsection{Expansion of the Transformed Hamiltonian}
We can now write the transformed Hamiltonian, $\exp[\mathcal{L}_\chi] H = \sum_{n=0}^\infty \frac{1}{n!}{\mathcal{L}^n_\chi}H$, in terms of $\Upsilon_{l}^n(H,\chi)$.
Since the lowest order terms appearing in $H$ are order $k'_\mathrm{min} = 2$, Equation \eqref{eq:lmin} becomes $l_\mathrm{min}^{(n)} = 2 +(k_\mathrm{min}-2)n$.
We have
\begin{equation}
 \exp[\mathcal{L}_\chi]H = 
    \sum_{n=0}^\infty
    \sum_{l=l_\mathrm{min}^{(n)}}^{\infty}
        \frac{1}{n!}\Upsilon^{n}_{l}(H,\chi) 
 = 
    \sum_{l=2}^\infty
    \sum_{n=0}^{N^{(l)}}
        \frac{1}{n!}\Upsilon^{n}_{l}(H,\chi) 
    := \sum_{l=2}^\infty (\Psi_{l} +[H_2,\chi_{l}]) 
 \label{eq:PsiDef}
\end{equation}
where we define $N^{(l)} = \floor{\frac{l-2}{k_\mathrm{min}-2}}$ and 
\begin{equation}
    \Psi_l = H_l + \sum_{k=k_\mathrm{min}}^{l-1} D_k H_{l+2-k} + \sum_{n=2}^{N^{(l)}}\frac{1}{n!}\Upsilon^{n}_l(H,\chi)~.
    \label{eq:PsiDef2}
\end{equation}
 The expression in Equation \eqref{eq:PsiDef2} defining $\Psi_l$ only contains terms, $\chi_k$, in the expansion of the generating function $\chi$ with $k<l$.
The terms of order $l$ gathered together in the sum in Equation \eqref{eq:PsiDef} are, by definition, the terms of order $l$ in the transformed Hamiltonian, $H'$. In other words, we have
\begin{equation}
 \Psi_l +[H_2,\chi_l] = H'_l~.
 \label{eq:homological_eqn}
\end{equation}
Since $\Psi_l$ contains only terms $\chi_k$ with $k<l$, we can iterate over $l$, solving for $H'_l$ and $\chi_l$ at each stage. Equation \eqref{eq:homological_eqn} is referred to as the homological equation.
\subsubsection{Solution of Homological Equation}
\label{sec:homological}
Recall that $\Psi_l$ is a homogeneous polynomial of degree $l$ in the complex variables $(x'_R,x'_z,\bar{x}'_R,\bar{x}'_z)$, i.e., a sum of terms 
$\propto x_R'^{k_R}x_z'^{k_z}\bar{x}_R'^{\bar{k}_R}\bar{x}_z'^{\bar{k}_z}$
with $k_R+k_z+\bar{k}_R+\bar{k}_z=l$.
Let $\langle\Psi_l\rangle$ denote the collection of terms in $\Psi_l$ with $k_R=\bar{k}_R$ and $k_z=\bar{k}_z$ and let 
\begin{equation}
    \{\Psi_l\} := \Psi_l-\langle\Psi_l\rangle = \sum_{i} C_ix_R'^{k_{R,i}}x_z'^{k_{z,i}}\bar{x}_R'^{\bar{k}_{R,i}}\bar{x}_z'^{\bar{k}_{z,i}}~.
\end{equation}  
Also note that
  \begin{equation}
      [H_2,\chi_{k}] = -\mathrm{i}\kappa\left(
        \bar{x}'_R\pd{\chi_{k}}{\bar{x}'_R}
        - 
        {x}'_R\pd{\chi_{k}}{{x}'_R}\right)
        -
        \mathrm{i}\nu\left(
        \bar{x}'_z\pd{\chi_{k}}{\bar{x}'_z}
        - 
        {x}'_z\pd{\chi_{k}}{{x}'_z}\right)~.
  \end{equation}
At each iteration, we solve Equation \eqref{eq:homological_eqn} by taking 
\begin{align}
    H'_l &= \langle\Psi_l\rangle\\
 \chi_l &= \sum_{i}
 \frac{\mathrm{i}C_i}
 {
    (k_{R,i} - \bar{k}_{R,i})\kappa 
    + 
    (k_{z,i} - \bar{k}_{z,i})\nu}
    x_R'^{k_{R,i}}
    x_z'^{k_{z,i}}
    \bar{x}_R'^{\bar{k}_{R,i}}
    \bar{x}_z'^{\bar{k}_{z,i}}~. \label{eq:chi_soln}
\end{align}
Equation \eqref{eq:chi_soln} assumes a non-resonance condition on $\kappa$ and $\nu$: we cannot have terms\footnote{Such resonant terms can instead be incorporated into $H'_{l}$ by simply omitting the corresponding terms from the sum in Equation \eqref{eq:chi_soln} defining $\chi_l$. The transformed Hamiltonian would, however, no longer be a function of actions alone.} with $(k_{R,i} - \bar{k}_{R,i})\kappa+(k_{z,i} - \bar{k}_{z,i})\nu = 0$.
\subsection{Azimuthal degree of freedom}
\label{sec:algorithm:azimuthal}
Thus far we have only considered the orbits of stars in the reduced phase space,  ignoring their azimuthal degree of freedom. Quasi-periodic orbits in the full 6 dimensional phase space will be characterized by a third dynamical frequency, $\Omega_\phi$,  in addition to the two frequencies, $\Omega_R=\frac{1}{x_R}\pd{}{\bar{x}_R}H'$ and $\Omega_z=\frac{1}{x_z}\pd{}{\bar{x}_z}H'$, readily obtained from the Birkhoff normalization procedure. 
Here we detail how this third frequency can be determined and used to define an  additional angle variable, $\theta_\phi$, once the Birkhoff normalization procedure has been applied to solve for orbits' dynamical evolution in the reduced phase space. 

The time derivative of the azimuthal angle, $\phi$ is given in terms of complex canonical variables by 
\begin{eqnarray}
     \frac{d}{dt}{\phi} = \frac{L}{R_C^2}
     \paren{1 + \frac{x_R+\bar{x}_R}{R_C\sqrt{2\kappa}}}^{-2}~.
    \label{eq:phi_dot}
\end{eqnarray}
For non-circular orbits, the right-hand side of Equation \eqref{eq:phi_dot} is not constant in time but will oscillate about a mean value that defines the azimuthal frequency, $\Omega_\phi$.  Introducing the canonical angle variable $\theta_\phi = \Omega_\phi t + \phi_0$ we can write the $\phi$ coordinate explicitly as
\begin{eqnarray}
    \phi = \theta_\phi + \rho_\phi(x'_R,x'_z,\bar{x}'_R,\bar{x}'_z,L)
    \label{eq:phi_of_aa}
\end{eqnarray}
where $\rho$ is a zero-mean $2\pi$-periodic function of $\theta_R=-\arg{x'_R}$ and $\theta_z=-\arg{x'_z}$.
, We can combine Equations \eqref{eq:phi_dot} and \eqref{eq:phi_of_aa} to express $\Omega_\phi$ and $\rho_\phi$ in terms of the  variables, $x'_R$ and $x'_z$, and their complex conjugates. Recalling the notation $\langle\cdot\rangle$ and $\{\cdot\}$ introduced above in Section \ref{sec:homological} to denote mean and oscillating terms of a Poisson series, the equations
\begin{eqnarray}
 \Omega_\phi &=& \frac{L}{R_C^2}
\left\langle\left(
    \exp[\mathcal{L}_\chi]
\sum_{m=0}^{\infty} \binom{-2}{m}
    \paren{
        \frac{x_R + \bar{x}_R}{R_C\sqrt{2\kappa}}
    }^{m}
\right)\right\rangle
\label{eq:Omega_phi_series}\\
 \dd{}{t}\rho_\phi &=&  \frac{L}{R_C^2}\left\{\left(
    \exp[\mathcal{L}_\chi]
\sum_{m=0}^{\infty} \binom{-2}{m}
    \paren{
        \frac{x_R + \bar{x}_R}{R_C\sqrt{2\kappa}}
    }^{m}
\right)\right\}~
\label{eq:rho_dot_def}
\end{eqnarray}
define $\Omega_\phi$ and the time derivative of $\rho_\phi$ as series in the variables $x'_R$ and $x'_z$ and their complex conjugates that can be expanded up to some maximum desired order. 
In order to derive an expression for $\rho_\phi$, we express the series on the right hand side of Equation \eqref{eq:rho_dot_def} as a sum of monomial terms
\begin{eqnarray}
    \left\{\left(
    \exp[\mathcal{L}_\chi]
\sum_{m=0}^{\infty} \binom{-2}{m}
    \paren{
        \frac{x_R + \bar{x}_R}{R_C\sqrt{2\kappa}}
    }^{m}
\right)\right\} 
= 
\sum_{i}A_ix_R'^{k_{R,i}}x_z'^{k_{z,i}}\bar{x}_R'^{\bar{k}_{R,i}}\bar{x}_z'^{\bar{k}_{z,i}}
\end{eqnarray} an then integrate with respect to time to obtain 
\begin{eqnarray}
\rho(x'_R,x'_z,\bar{x}'_R,\bar{x}'_z,L) = \mathrm{i}\frac{L}{R_C^2}\sum_{i}
 \frac{
 A_i
x_R'^{k_{R,i}}x_z'^{k_{z,i}}\bar{x}_R'^{\bar{k}_{R,i}}\bar{x}_z'^{\bar{k}_{z,i}} }
 {
    (k_{R,i} - \bar{k}_{R,i})\kappa 
    + 
    (k_{z,i} - \bar{k}_{z,i})\nu
}
    ~.
\end{eqnarray}
It follows immediately from Hamilton's equations that 
$J_\phi = L$ serves as canonical action coordinate conjugate to the angle $\theta_\phi$.
We therefore have established a complete canonical transformation from the original variables $(R,z,\phi,p_R,p_z,L)$ of the 6 dimnensional phase space into to a complete set AA variables, $(\theta_R,\theta_z,\theta_\phi,J_R,J_z,J_\phi)$ where $J_i =|x'_i|^2$ and $\theta_{i} = -\arg x'_i$ for the subscript $i\in \{z,R\}$.

\subsection{Numerical Implementation}
The  Birkhoff normalization algorithm  described above is implemented in the \texttt{poisson\_series} module of the open-source Python package \texttt{celmech} \citep{hadden_celmech_2022}. This module provides routines for representing and manipulating so-called ``Poisson series" \citep[][]{Brumberg1995}. These are series in $N$ complex canonical variables $x_i$ and their complex conjugates, $\bar{x}_i$ along with $M$ real action-angle pairs, $(P_i,Q_i)$, that are given as sums of individual monomial terms of the form
\begin{equation}
     {\cal M}(k,\bar{k},p,q) = \prod_{i=1}^{N} x_i^{k_i} \bar{x}_i^{\bar{k}_i } \prod_{j=1}^{M} P_j^{p_j}\exp[i q_j Q_j]~.
     \label{eq:mon}
\end{equation}
The Poisson bracket of any pair of Poisson series is another Poisson series, as can be readily seen by considering the Poisson bracket of two monomial terms. Defining the length-$N$ vectors $o^N_i$ such that $[o^N_i]_j = \delta_{ij}$ where $\delta_{ij}$ is the Kroenecker delta, the Poisson bracket of two monomial terms is then a Poisson series consisting of (at most) $N+M$ new monomial terms and given by
\begin{multline}
[{\cal M}(k,\bar{k},p,q),{\cal M}(l,\bar{l},r,s)]
 = \mathrm{i}\sum_{i=1}^N (\bar{k}_i{l}_i-k_i\bar{l}_i){\cal M}(k+l-o^N_i,\bar{k}+\bar{l}-o^N_i,p+r,q+s)
 +
 \\
 \mathrm{i}\sum_{j=1}^{M}(q_jr_j-s_jp_j){\cal M}(k+l,\bar{k}+\bar{l},p+r-o^M_j,q+s)~.
\end{multline}
The \texttt{poisson\_series} module defines the \texttt{PoissonSeries} class to store sums of monomial terms of the form given in Equation \eqref{eq:mon} multiplied by numerical coefficients. It also defines a number of routines for manipulating these series, including performing scalar multiplication, multiplying or adding pairs of series, and calculating Poisson brackets. Note that the Poisson series used in this paper involve only complex canonical variables (i.e., $M=0$). 

The \texttt{poisson\_series} module additionally defines routines for carrying out the Birkhoff normalization procedure and evaluating finite-order approximations of  exponential operators, $\exp[\mathcal{L}_\chi]$, applied to series. These routines require grouping terms in the expansions of various functions, such as the Hamiltonian, $H$, and the generating function, $\chi$, by their order in powers of the complex canonical variables. In the framework of the \texttt{poisson\_series} module, this grouping is done by using \texttt{Python} dictionary objects to associate \texttt{PoissonSeries} objects representing terms in an expansion of a given order to key values that indicate their order. Example Jupyter notebooks illustrating the use of \texttt{celmech}'s \texttt{poisson\_series} module to carry out the Birkhoff normalization procedure are available at \href{https://github.com/shadden/AA_variables_via_Birkhoff_Normalization}{github.com/shadden/AA\_variables\_via\_Birkhoff\_Normalization}.
\subsection{Padé Approximants}
\label{sec:pade}
Since the Birkhoff normalization procedure is based upon a Taylor expansion of the effective potential, the method can only be expected to provide accurate results for orbits that remain within regions of phase space where this expansion is convergent. In practice, this can  make the procedure inapplicable to orbits that experience significant excursions in the vertical direction. For example, the Miyamoto-Nagai potential, 
\begin{equation}
    \Phi(R,z) = -\frac{1}{\sqrt{R^2 + (a+\sqrt{z^2+b^2})^2}}~
    \label{eq:MN_potential}
\end{equation}
 with parameters $a$ and $b$, is a commonly adopted as a model for the galactic disk's contribution to the gravitational potential \citep{MiyamotoNagai1975}. The parameter $b$ effectively sets the thickness of the galactic disk and the branch points of Equation \eqref{eq:MN_potential} occurring in the complex plane at $z=\pm ib$ restrict the convergence Taylor series expansions to  $|z|<b$.
 
 Accurate local approximations of functions can sometimes be extended beyond the domain of convergence of their Taylor series with the use of Padé approximants.  The $(n,m)$ Padé approximant of a function is the rational function with numerator of degree $m$ and denominator of degree $n$ that agrees with the Taylor series expansion of the function up to degree $x^{m+n}$ \citep[e.g.,][]{teukolsky1992numerical}.  If the first $n+m$ terms of a function's Taylor series are $\sum_{k=0}^{n+m} c_kx^k$ then the coefficients of an $(n,m)$ Padé approximant are readily obtained by multiplying both left and right hand sides of 
 \begin{eqnarray}
    \sum_{k=0}^{n+m} c_kx^k  = \frac{a_0+a_1 x+...a_{m}x^{m}}{1+b_1x + ... +b_nx^n}\mod{x^{n+m+1}}
    \label{eq:pade_equiv}
\end{eqnarray}
by $1+b_1x + ... +b_nx^n$. The coefficients, $b_i$,  satisfy the linear equation
\begin{eqnarray}
    \begin{pmatrix}
    c_{n+m-1} & \hdots  & c_{m} 
    \\
    \vdots    & \ddots      &  \vdots 
    \\
    c_{m}    & \hdots  & c_{m-n+1}
    \end{pmatrix}
    \cdot
    \begin{pmatrix}
    b_1\\
    \vdots\\
    b_n
    \end{pmatrix}
    = 
    -
    \begin{pmatrix}
    c_{n+m}\\
    \vdots\\
    c_{m+1}
    \end{pmatrix}.
\end{eqnarray}
and the coefficients, $a_i$, are then given by $a_i = \sum_{k=0}^{n}c_{i-k}b_k$ where $b_0 :=1$ and $c_k$ is taken to be $0$ for $k<0$.
 
 Below in Section \ref{sec:applications},  we show that constructing Padé approximants in the variable $I_z=|x_z|^2$, often improves the accuracy of the Birkhoff normalization procedure. The Padé approximants involve coefficients, $b_i$ and $a_i$, that are functions of $x_R,~\bar{x}_R$ and $\phi_z=-\arg(x_z)$. These coefficients are determined by evaluating finite-order truncations of the series
 $x'_R = \exp[-\mathcal{L}_\chi]x_R$ and $x'_z = \exp[-\mathcal{L}_\chi]x_z$ and then grouping terms by powers of $I_z$. Note that the generating function, $\chi$, is developed as a series  in powers of $x_z$ and $\bar{x}_z$, which are both proportional to $\sqrt{I_z}$. However, it is straightforward to show that for any $\Phi(R,z)$ like Equation \eqref{eq:MN_potential} that is an even function of $z$, only integer powers of $I_z$ will appear in $\chi$ as well as finite-order truncations of $\exp[-\mathcal{L}_\chi]x_R$.  Thus, these finite-order truncations are polynomials in $I_z$ (rather than $\sqrt{I_z}$) and can be used to obtain Padé approximants in $I_z$. Additionally, finite-order truncations of $\exp[-\mathcal{L}_\chi]x_z$ yield expressions that can be written as $x_z P(I_z)$ where $P$ is a polynomial in $I_z$ with coefficients that depend on $x_R,\bar{x}_R,$ and $\phi_z$. Padé approximants are derived from the polynomial $P$ and used to calculate $\exp[-\mathcal{L}_\chi]x_z$ in Section \ref{sec:applications}.
\section{Tests}
\label{sec:applications}
In this section  we apply the Birkhoff normalization procedure to orbits in the Miyamoto-Nagai potential in Equation \eqref{eq:MN_potential}. The parameters $a = 3$, $b = 0.3$, and $L = 3$ are chosen, with a corresponding circular orbit radius of $R_C(L)\approx 10.4$. With distance units taken to be kpc, these parameters approximately match the parameters adopted for the Miyamoto-Nagai component of the Milky Way potential model, \texttt{MWPotential2014}, implemented in the \texttt{galpy} Python package \citep{Bovy2015}. The Birkhoff normalization procedure is carried out by developing the generating function, $\chi$, up to 10th order in complex canonical variables. The \texttt{sympy} symbolic mathematics package \citep{sympy} is used to calculate the requisite partial derivatives of the effective potential.

  We also compute actions and frequencies using the Stäckel approximation method introduced by \citet{binney_actions_2012}. These calulations are done using the \texttt{actionAngle.actionAngleStäckel} function implemented in \texttt{galpy} \citep{BovyRix2013,Bovy2015}. This function requires as input an estimated focal length parameter, $\Delta$, in addition to a potential and orbital initial conditions. For all orbits considered, The \texttt{galpy} function \texttt{actionAngle.estimateDeltaStäckel} is used to estimate the focal length  based on partial derivatives of the potential evaluated at $(R,z) = (R_C,0)$ following \citet{Sanders2012}.

\begin{figure}
    \centering
    \includegraphics[width=1\textwidth]{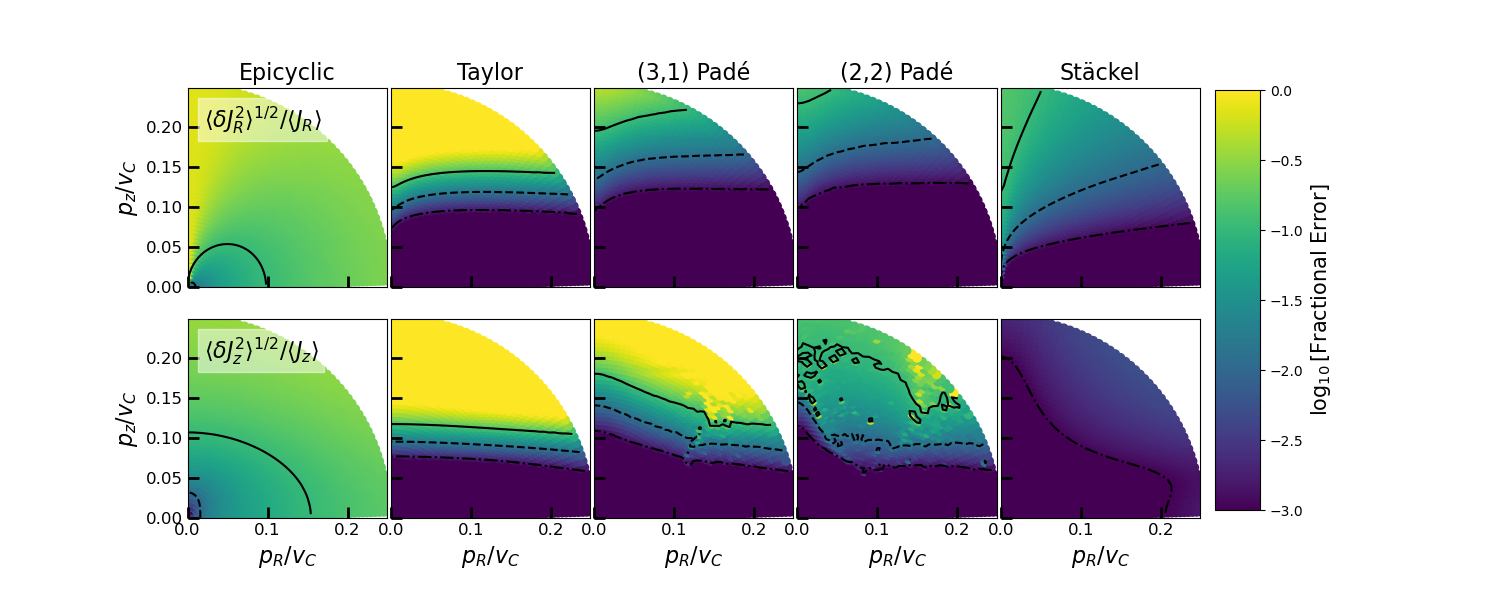}
    \caption{Color maps showing r.m.s. fractional variations in  estimated values of actions on a grid of orbits in a Miyamoto-Nagai potential (Equation \ref{eq:MN_potential}) with parameters $a=3$ and $b=0.3$. 
    Actions are computed along orbits starting with $(L,R,z) = (3,R_C,0)$ and initial velocity values indicated by the axes, which are plotted in units of the circular velocity, $v_C = L/R_C$.
    The top row shows variations in the radial action, $J_R$, while the bottom row shows variations in the vertical action, $J_z$. 
    Columns are labeled according to the method used to compute action values. See text for additional details. Dash-dot, dashed, and solid contours mark fractional error levels of $0.1\%, 1\%$ and $10\%$, respectively.
    }
    \label{fig:action_rms_error}
\end{figure}
Figure \ref{fig:action_rms_error} illustrates the accuracy of actions determined by Birkhoff normalization and Stäckel approximation across a range of initial conditions. Orbits are initialized at $(R,z) = (R_C,0)$ over the plotted range of $(p_R,p_z)$ values and integrated for 10 times the radial epicyclic period (i.e., $T = 10\times (2\pi/\kappa)$) with 512 outputs generated at equally-spaced intervals along the orbits. Transformed complex canonical variable, $x'_i$,  are computed as functions of the un-transformed variables along a given orbit using various approximations to the  series
\begin{eqnarray}
\label{eq:xprime_of_x}
    x'_i(x_R,x_z,\bar{x}_R,\bar{x}_z) = (\exp[-\mathcal{L}_\chi]x_i)(x_R,x_z,\bar{x}_R,\bar{x}_z)
\end{eqnarray} 
where  the subscript $i$ denotes $R$ or $z$. Action values are then computed from the transformed complex canonical variable as $J_i = |x'_i|^2$. The root-mean-square (r.m.s.) fractional variations in these action values computed along each orbit are recorded in the color scale. 

The leftmost column of Figure \ref{fig:action_rms_error} show the r.m.s. variations in actions computed using the epicyclic approximation, where Equation \eqref{eq:xprime_of_x} is approximated simply as $x_i'\approx x_i$. 
The variations in actions computed via this approximation provide a measure of how significantly orbits are affected by the anharmonicity of the potential. 
The second column of Figure \ref{fig:action_rms_error} shows the r.m.s. variations for actions computed by expanding the right hand side of Equation \eqref{eq:xprime_of_x} as a Taylor series. With $\chi$ determined up to 10th order in the canonical variables,  the terms of these series are determined up to 9th order in the complex canonical variables. The third and fourth  columns show r.m.s. variations for actions computed by constructing $(3,1)$ and $(2,2)$ Padé approximants in $I_z$ as described in Section \ref{sec:pade}.  The final columns the r.m.s. variations for actions computed using the Stäckel approximation. 

The Taylor series approximations of actions in Figure \ref{fig:action_rms_error}  provide good agreement with the numerical results for  mid-plane vertical  velocity values  $p_z\lesssim 0.1v_C$, where $v_C = L/R_C$ is the circular orbital velocity. The poor agreement for greater vertical velocities is attributable to the fact that orbits with greater vertical velocities reach heights $z>b$, beyond the radius of convergence of the Taylor series expansion of the potential in Equation \eqref{eq:MN_potential}. An estimate of the maximum vertical height, $z_\mathrm{max}$, achieved by an orbit with a given vertical velocity, $p_{z,0}$, in the midplane can be obtained by neglecting any motion in the radial direction and using conservation of energy to  equate $\frac{1}{2}p_{z,0} \approx \Phi_{\mathrm{eff}}(R_C,z_\mathrm{max}) - \Phi_{\mathrm{eff}}(R_C,0)$. Taking $z_\mathrm{max} = b$, the corresponding midplane velocity is $p_{z,0}\approx0.09 v_C$, closely matching the value of mid-plane velocity above which the Taylor series approximations fail in Figure \ref{fig:action_rms_error}.  The Padé approximants clearly extend the range over which the Birkhoff normalization procedure accurately predicts actions, achieving r.m.s. variations in radial actions of $\lesssim10\%$ for initial vertical velocities $p_z\lesssim0.2v_C$. The (2,2) Padé approximant is slightly more accurate than the (3,1) approximant. Performance of the algorithm for vertical action determination is slightly worse, with $\lesssim10\%$ accuracy restricted to initial vertical velocities $p_z\lesssim0.15v_C$.

\begin{figure}
    \centering
    \includegraphics[width=1\textwidth]{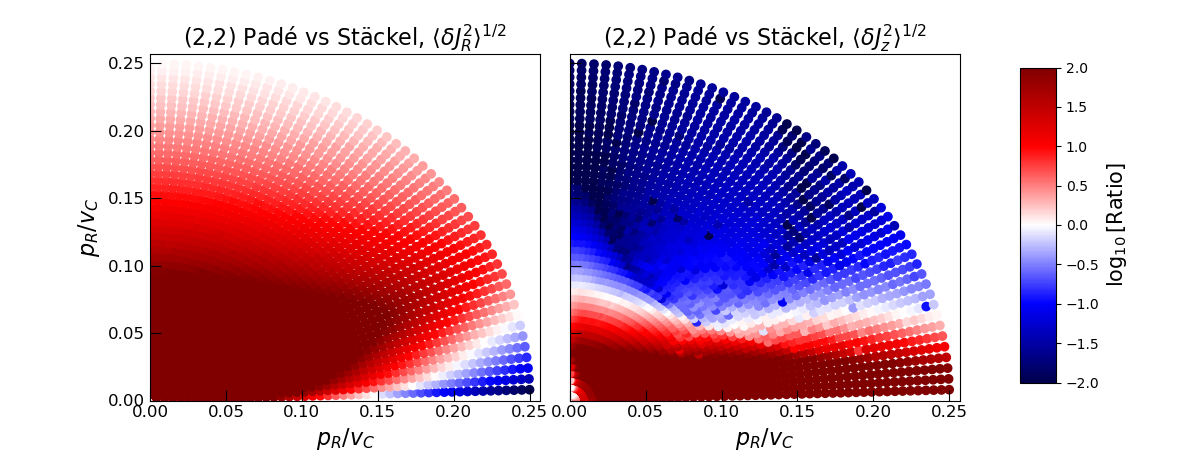}
    \caption{Comparison of the accuracy of actions determined using Birkhoff normalization and (2,2) Padé approximants versus the Stäckel approximation. The color scale indicates the r.m.s. variation in actions computed using  the Stäckel approximation divided by r.m.s. variations computed via Birkhoff normalization so that red colors indicate orbits where Birkhoff normalization yields better accuracy while blue colors indicate orbits for which the Stäckel approximation is more accurate.
    The grid of orbits is the same as shown in Figure \ref{fig:action_rms_error}.
    }
    \label{fig:actions_compare}
\end{figure}

Figure \ref{fig:actions_compare}  compares the accuracy of action determinations made using Birkhoff normalization with (2,2) Padé approximants against the Stäckel approximation. Birkhoff normalization provides a more accurate determination of radial actions over the majority of the plotted range. Vertical actions are also more accurately determined for $p_z\lesssim 0.1 v_C$.

\begin{figure}
    \centering
    \includegraphics[width=1\textwidth]{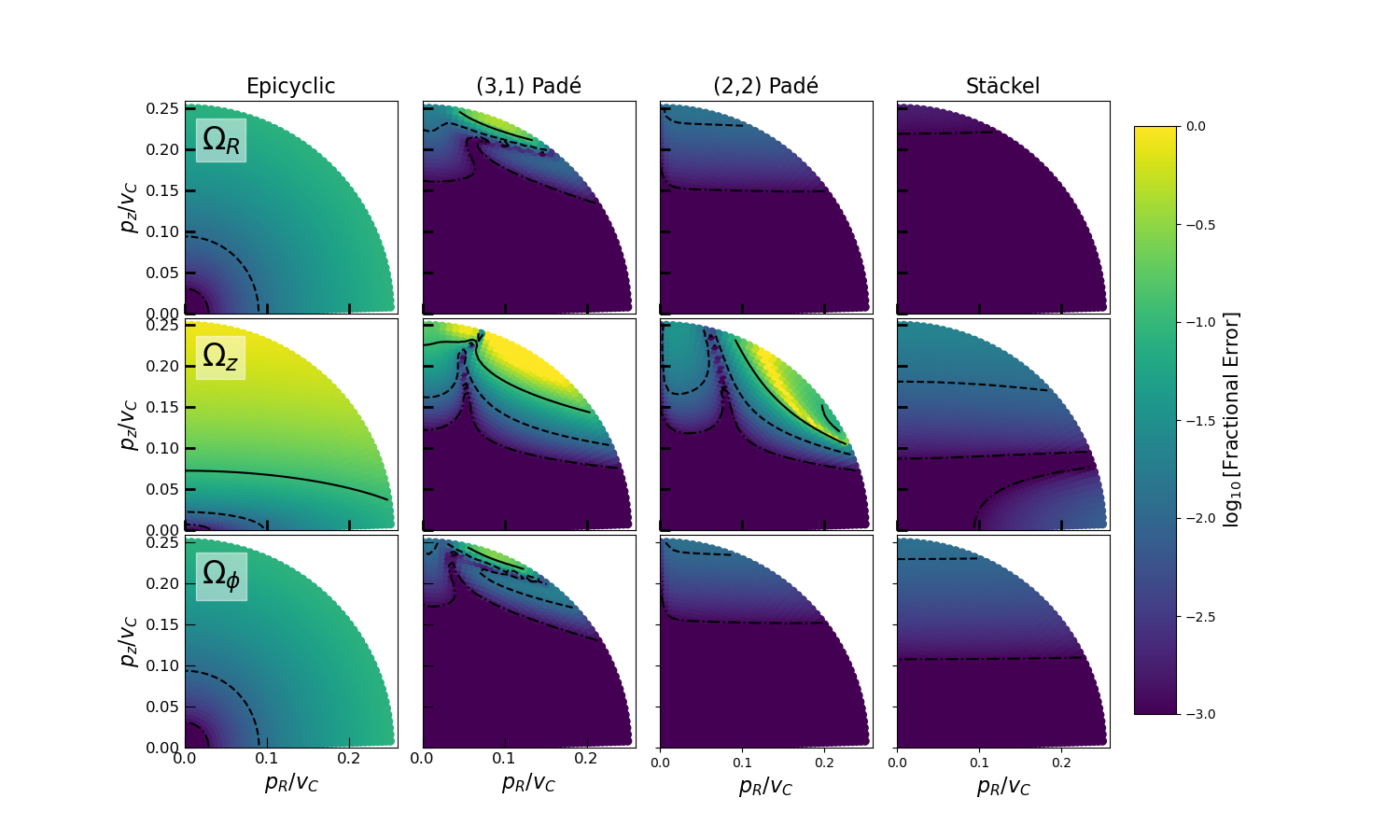}
    \caption{
    Color maps showing fractional errors in the estimated values of radial, vertical, and azimuthal frequencies on a grid of orbits in the Miyamoto-Nagai potential (Equation \eqref{eq:MN_potential}) with parameters $a=3$ and $b=0.3$. 
    Frequencies are computed for orbits starting with $(L,R,z) = (3,R_C,0)$ and initial velocity values indicated by the axes, which are plotted in units of the circular velocity, $v_C = L/R_C$.
    The top, middle, and bottom rows shows fractional errors in the radial ($\Omega_R$), vertical ($\Omega_z$), and azimuthal ($\Omega_\phi$) frequencies, respectively. 
    Columns are labeled according to the method used to compute the frequency values. See text for details. Dash-dot, dashed, and solid contours mark fractional error levels of $0.1\%, 1\%$ and $10\%$, respectively.}
    \label{fig:freq_error}
\end{figure}

Figure \ref{fig:freq_error} illustrates the accuracy of the determination of the dynamical frequencies $\Omega_R$, $\Omega_z$, and $\Omega_\phi$  using various methods. Exact values of frequencies are determined numerically using the frequency modified Fourier transform algorithm of \cite{fmft1996}, implemented in the \texttt{celmech} package as the function \texttt{miscellaneous.frequency\_modified\_fourier\_transform}. The method was applied to the time series of complex canonical variables, $x_R$ and $x_z$, for each orbit integration and the frequencies of the largest Fourier components determined for each variable were identified as $-\Omega_R$ and $-\Omega_z$, respectively. Values of $\Omega_\phi$ were determined by applying the same method to time series of $e^{\mathrm{i}\phi}$. The first column of Figure \ref{fig:freq_error} records the fractional errors $\kappa/\Omega_R - 1$, $\nu/\Omega_z - 1$, and $R_C^{-2}L/\Omega_\phi-1$ incurred by adopting the epicyclic approximation for the dynamical frequencies.  These panels illustrate that the frequencies $\Omega_R$ and $\Omega_\phi$ vary gradually with both initial radial and vertical velocities while the frequency $\Omega_z$ is a strong function of the initial vertical velocity.

The second  and third columns of Figure \ref{fig:freq_error} record fractional errors incurred by using Birkhoff normalization and Padé approximants to determine frequencies. Frequencies were determined as follows. First, values of complex canonincal variables were assigned based on orbits' initial conditions as $x_R = \mathrm{i}\sqrt{\frac{1}{2\kappa}}p_R$ and $x_z = \mathrm{i}\sqrt{\frac{1}{2\nu}}p_z$ (see Equations \ref{eq:xR_def} and \ref{eq:xz_def}). Next, values of transformed complex canonical variables, $x'_R$ and $x'_z$, were computed using the  Padé approximation method described in Section \ref{sec:pade}.  Then, series expressions for frequencies, $\Omega_R(J_R,J_z) = \partial_{J_R}H'$ and $\Omega_z(J_R,J_z) = \partial_{J_z}H'$, derived from the transformed Hamiltonian, $H'$, and the series expression for $\Omega_\phi$ given by Equation \eqref{eq:Omega_phi_series} were computed. Finally, these series expressions were used to construct a second set of (3,1) and (2,2) Padé approximants, this time in $J_z=|x'_z|^2$, and evaluated. Comparing Figures \ref{fig:action_rms_error} and \ref{fig:freq_error}, we see that dynamical frequencies are accurately recovered in essentially the same regions of phase space that Birkhoff normalization and Padé approximants provide accurate action variables. The fourth column of Figure \ref{fig:freq_error} shows frequency errors incurred using the Stäckel approximation for comparison. 

\section{Summary \& Discussion}
\label{sec:summary}
We have shown how Birkhoff normalization, a technique from Hamiltonian perturbation theory, can be algorithmically implemented and applied to derive AA variables for stellar orbits in axisymmetric galactic potentials.  One significant advantage of this method  is that it provides explicit expressions for actions and angle variables as functions of positions and velocities and vice versa. 
It similarly yields an explicit expression for the Hamiltonian as a function of the action variables. The only input required by the method is the partial derivatives of the effective potential up to the maximum desired order of the series expansions. 

Since  Birkhoff normalization relies on expansion of the effective potential about its value for circular, planar  orbits, its range of applicability is nominally limited by the radius of convergence of this expansion. However, as illustrated by the tests in Section \ref{sec:applications}, the series expressions generated by the procedure can be used to construct Padé approximants and extend the range over which it provides accurate results.
The tests presented in Section \ref{sec:applications} showed that Birkhoff normalization provides comparable or better accuracy than the popular Stäckel approximation, at least for quasi-planar orbits. 
 
 The expressions provided by Birkhoff normalization take the form of polynomials involving the complex canonical variables introduced in Section \ref{sec:cc_vars}. 
 This makes evaluating both forward and inverse transformations between position-velocity data in AA variables exceptionally computationally efficient.
 This computational efficiency represents another potential advantage of the method over the Stäckel approximation, which requires evaluation of integrals via numerical quadrature. 
The most computationally expensive parts of the Birkhoff normalization algorithm will generally be the initial calculation of potential derivatives and construction of series.
However, these calculations represent a fixed cost and need only be performed once, after which the Birkhoff normalization procedure can be applied to as many orbits as desired. 
For the tests presented in Section \ref{sec:applications}, computing partial derivatives of the Miyamoto-Nagai potential up to 10th order via symbolic means required $\sim 20~\mathrm{s}$ on a 2.5 GHz Intel i7 processor. 
The initial construction of the series expression for $\chi$ up to 10th order in complex canonical variables took  $\sim 3~\mathrm{s}$ and calculations of additional series expressions approximating the exponential operators $\exp[\pm\mathcal{L}_\chi]$ applied to various functions, such as Equation \eqref{eq:xprime_of_x}, took approximately $\sim 1~\mathrm{s}$ each.
A potential downside to the Birkhoff normalization method, in terms of computational cost, is that the effective potential partial derivatives and series must nominally be recalculated for each value of angular momentum considered. 
A simple strategy for applying the method to orbits with a range of angular momenta would be to carry out series constructions on a 1 dimensional grid of angular momenta and simply interpolate results. 
Alternatively, the construction of various series could be carried out symbolically with the aid of a computer algebra system. 
We leave the development of such methods to future work as the choice of optimal strategy is likely to be application-specific.

We conclude by mentioning a few potential future applications of the Birkhoff normalization procedure presented in this paper.
First, analytic expressions furnished by the Birkhoff normalization procedure could make it well-suited for problems requiring the representation of equilibrium distribution functions, which by Jean's theorem must depend only on integrals of motion, in terms of spatial positions and velocities.
Additionally, the explicit expression for the Hamiltonian in terms of action variables provided by the Birkhoff normalization procedure can be used as a starting point for using perturbation theory to understand the non-equilibrium signatures of dynamical perturbations like those induced by a stellar bar \citep[e.g.,][]{binney_orbital_2018} or satellite galaxy \citep[e.g.,][]{banik_comprehensive_2022}.  
Finally, even though the Birkhoff normalization procedure may not supply accurate AA variables for orbits with large vertical actions, it could provide useful initial `toy' AA variables that serve as the starting point of the torus mapping method of \citet{mcgill_torus_1990}.
\section{Acknowledgments} 
\label{sec:acknowledgments}
I thank Rimpei Chiba and Neige Frankel for helpful discussions. I thank Scott Tremaine for suggesting the use of Padé approximants. Simulations were performed on the Sunnyvale cluster at the Canadian Institute for Theoretical Astrophysics. I acknowledge support by the Natural Sciences and Engineering Research Council of Canada (NSERC), funding references CITA 490888-16 and RGPIN-2020-03885. This project was developed in part at the Gaia Hike, a workshop hosted by the University of British Columbia and the Canadian Institute for Theoretical Astrophysics in 2022 June.
\bibliographystyle{aasjournal}
\bibliography{zotero_lib,more_refs}
\end{document}